# IP Video Conferencing: A Tutorial

Roman Sorokin, Jean-Louis Rougier

**Abstract** Video conferencing is a well-established area of communications, which have been studied for decades. Recently this area has received a new impulse due to significantly increased bandwidth of Local and Wide area networks, appearance of low-priced video equipment and development of web based media technologies. This paper presents the main techniques behind the modern IP-based videoconferencing services, with a particular focus on codecs, network protocols, architectures and standardization efforts. Questions of security and topologies are also tackled. A description of a typical video conference scenario is provided, demonstrating how the technologies, responsible for different conference aspects, are working together. Traditional industrial disposition as well as modern innovative approaches are both addressed. Current industry trends are highlighted in respect to the topics, described in the tutorial. Legacy analog/digital technologies, together with the gateways between the traditional and the IP videoconferencing systems, are not considered.

**Keywords** Video Conferencing, codec, SVC, MCU, SIP, RTP

Roman Sorokin

ALE International, Colombes, France

e-mail: roman.sorokin@al-enterprise.com

Jean-Louis Rougier

Télécom ParisTech, Paris, France

e-mail: rougier@telecom-paristech.fr

# 1 Introduction

Video conferencing is a two-way interactive communication, delivered over networks of different nature, which allows people from several locations to participate in a meeting.

Conference participants use video conferencing endpoints of different types. Generally a video conference endpoint has a camera and a microphone. The video stream, generated by the camera, and the audio stream, coming from the microphone, are both compressed and sent to the network interface. Some additional info like instant messages, the shared screen or a document can be also exchanged between participants.

IP video conferencing, considered in this tutorial, is based on the TCP/IP technology as a transport network for all these flows. In the past, specially designed analog lines and digital telephonic lines (ISDN) had been employed for that purpose. IP started to be used in the 1990s and has become the prominent vehicle for video conferencing since then.

Today IP video conferencing is a well known and widely used service. However, most users might not realize that it has a notably complex architecture, involving a wide range of technologies. This tutorial aims at providing an overview of possible architectures and technologies, involved in the realization of videoconferencing services. Any reader with basic information technology and networking knowledge should be able to read this paper.

The rest of the paper is organized as follows. An example of a possible video conference functional architecture is presented in chapter 2. The notion and technologies of video coding is discussed in chapter 3. Description of video processing is provided in chapter 4. Protocols, utilized in video conferencing, are introduced in chapter 5. Types of media servers are described in chapter 6 and types of clients in chapter 7. Video conferencing topologies are presented in chapter 8. Current industry trends are drawn in chapter 9.

# 2 Functional architecture example

There exist two fundamental means to set up videoconference calls between participants. Basic conference functions can be offered in peer-to-peer mode, in which all the participants are connected directly with each other (see Fig. 1).

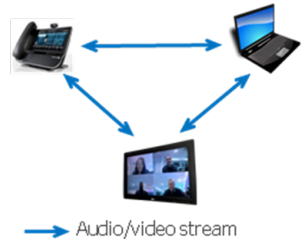

Fig. 1. Peer-to-peer video conference

Conferences, which provide more services, such as central management of participants or conference recording, generally make use of a central point of control ("Middlebox") in order to implement these additional services (see Fig. 2).

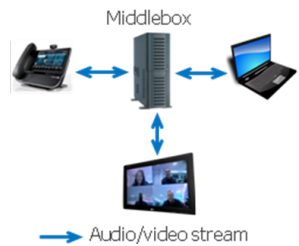

Fig. 2. Video conference with a middlebox

In this section, we present an example of a possible functional architecture of a conferencing solution, using several dedicated servers (defined hereafter), as depicted in Fig. 3. These servers play a role of the "middlebox" in the centralized conferencing architecture. Such architecture is typical for advanced video conferences in enterprises.

## 2.1 Functional elements

**Endpoint:** A software application or dedicated hardware equipment, which allows a user to participate in a conference. It consists of the following elements:

- Equipment for capturing and rendering both audio and video: a screen, a microphone and a loudspeaker or headphones
- Audio/video coder and decoder, in order to limit the throughput of streams sent on the network
- A signaling protocol stack, which is responsible for the registering the user in the conferencing system, joining or leaving a conference, negotiation of media formats, etc.
- A media transport protocol stack, which delivers encoded media over the network between endpoints and the middlebox

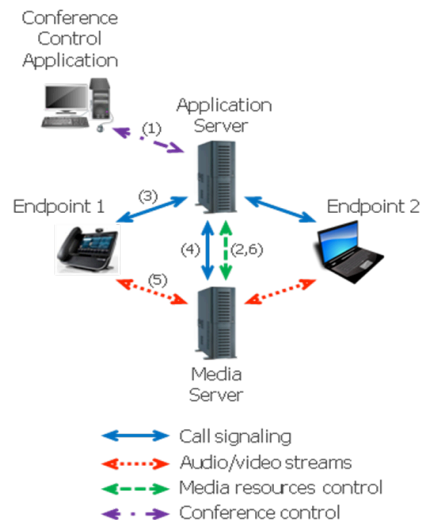

Fig. 3. An example of a video conference functional architecture

**Conference Control Application:** A Graphic User Interface application, which allows the conference leader to fulfill different operations on the conference, such as reserving media resources for a future conference, inviting new participants or removing existing participants. Web technologies are often used for this type of applications, which can also be integrated with the endpoint software.

**Media Server:** Software component or hardware appliance, which comprises resources for media processing, like:

- Audio mixing, that allows the voices of conference participants to be mixed into one stream, that can be sent to all the participants
- Video mixing, that allows the images of several participants to be shown simultaneously on the screen (see Fig. 4)

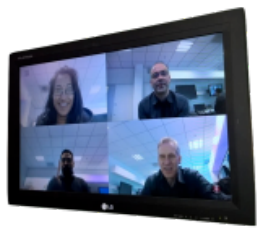

Fig. 4. Multi image video conference endpoint

Basic media processing facilities, like media mixing, can be integrated into endpoints and can thus be used in peer-to-peer mode. The use of a centralized Media Server gives some advantages like:

- Media trans-coding if media is encoded by different endpoints in incompatible formats
- Generating additional media streams based on the conference ones, for example for recording

**Application server:** A software component or hardware appliance, which plays the central management role between other functional parts of the video conferencing solution (Endpoints, Conference Control Application and Media Server). Its functionality encompasses:

- Localization of Endpoints (Endpoints register in the Application Server so their network locations are known) and management of Call Signaling sessions

- Conference management: processing the requests of the conference leader made with the Conference Control Application (inviting/removing users, etc.) and translating them to Call Signaling session commands towards respective Endpoints

- Media Server management: based on the logic of the conference Application Server, the Media Server applies different media treatment to the conference participants, like playing a voice message to the participants, recording the session, etc.

Having management functions centralized allows the conference to continue smoothly even while some Endpoints leave the conference --- which is hardly possible in the case when management logic resides on one of the Endpoints. Furthermore, centralized management facilitates integration with different enterprise software, like corporate directory with information about employees, shared calendars, etc.

Application and Media Servers can be combined in one box with specifically selected hardware optimized for delivering high quality audio/video experience.

### 2.2 Workflow example

The dynamic view of the architecture, presented in Fig. 3, is demonstrated with the scenario below. The technologies and the protocols, used for this demonstration, are quite typical for videoconferencing, deployed in modern enterprises: SIP (presented in 5.2) is used for call signaling, RTP (explained in 5.1) is used for streaming media on the network and MSML (highlighted in 5.3) for media resources control. The scenario, depicted in Fig. 3. follows several steps:

1. The conference leader creates a conference using the Conference Control Application (step 1).

2. The Application Server sends a MSML command to the Media Server with instructions on how the conference must be configured and which resources are needed (step 2).

3. At this stage, Endpoint1 wants to join the conference. It sends the SIP request to the Application Server to join the video conference (step 3). The request includes information about the media session parameters.

4. The Application Server forwards the request towards the Media Server (step 4), which in turn sends a SIP answer to Endpoint1, with its own media session parameters.

5. A connection between Endpoint 1 and the Media Server is now opened (step 5). This is a direct RTP session as the Application Server does not process media streams. The connection between Endpoint1 and the Media Server is operational but nothing is transported yet, as we need to attach this session to a source of media in the Media Server (for instance to a video mixing function).

6. The Application Server sends a MSML command, which allows the connection of Endpoint1 to be attached to the videoconference session (step 6). From this moment, the media flows generated by Endpoint1 will be mixed with the streams of other participants.

If another endpoint (Endpoint2) joins the conference, the procedure will be exactly the same as described in steps 3-6 above.

## 3 Video coding

### 3.1 Why video coding

End user's device for capturing video (i.e. web camera integrated into laptop) produces raw (uncompressed) digital video stream. Video processing (i.e. video mixer in media server) and video rendering (i.e. video conferencing endpoint) also require uncompressed digital video streams. However, raw digital video streams are usually too heavy (i.e. they consume too much bandwidth) to be sent through the network, so they should be compressed.

Video encoding is a process of converting raw digital video to a compressed format, video decoding is the opposite process. A hardware or software component that fulfills encoding and decoding is called "codec" (which is a concatenation of "coder" and "decoder") [1].

The format of the compressed streams normally conforms to some video compression standards. The standards typically define lossy compression, meaning that the compressed stream loses some of the original information present in the raw stream. As a result, compressed/decompressed streams have lower quality than the original ones.

## 3.2 Types of video coding

The codecs differ by the quantity of the data, needed to transport the video stream (which is called "bitrate"), the complexity of the encoding and decoding algorithms, robustness to data losses and errors, which occur when the stream traverses the network, end-to-end delay, and a lot of other parameters.

User endpoints vary in their capabilities to accept and process video streams. These differences can be explained by:

- Different bandwidth capabilities
- Different decoding complexity and power constraints

For example, a specialized hardware based video conferencing endpoint, which is normally installed in a meeting room, is typically able to process high quality video streams. However, a participant using a smart phone is only able to process low quality streams using small bitrates. Generally, this problem is resolved by a "Transforming middlebox", which can adjust streams to the recipients' needs (see Fig. 5).

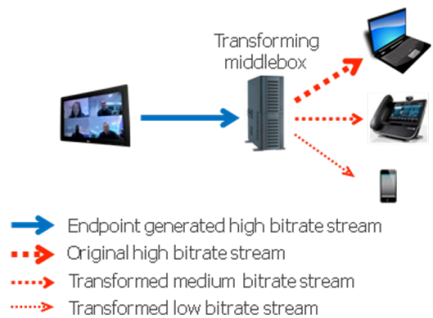

Fig. 5. Logic of transforming middlebox

Scalable Video Coding (SVC) is another approach, allowing different types of devices to participate in the same video conference. With SVC, a high-quality video stream is divided into several layers of quality. For instance in Fig 6, the three layers are sent on the network. The mobile terminal (with poor network reception) will only receive the base layer, which corresponds to the lowest video quality. The other terminals, which can benefit from a better network throughput and/or CPU power, can receive additional layers (on top of the base layer) in order to get a better video quality.

The advantage of this technique is that processing at the SVC middlebox is extremely light, as the middlebox just needs to filter the different built-in layers, and processing of the content of the video stream is not required.

One of the following methods can be used to build the different layers:

- Temporal scalability (frame rate): a low frame rate is used for the base layer, while additional frames are added on advanced layers, providing more fluidity to the video.
- Spatial scalability (picture size): the base layer has a low image resolution, while advanced layers add additional pixels increasing it.
- Quality scalability (coding quality): the coding quality corresponds to the number of bits associated with each pixel. The base layer is coded with the low coding quality, advanced layers with the better one.

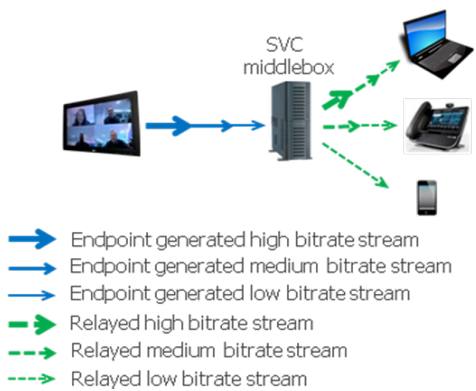

- 
- Fig. 6. Logic of SVC middlebox

It is also possible to combine several of the above scalability techniques.

The flexibility of SVC comes at a price, since the layer encoding adds a bandwidth overhead of roughly 10% - 20%, as compared with a non SVC stream of the same quality.

Unfortunately, SVC technique is not currently fully standardized, so implementations of different vendors are not compatible with each other, except for the base low bitrate layer, which is coded as standard stream (and can thus be decoded by decoders which understand only standard coding).

In order to overcome this obstacle, another method called Simulcast Video Coding was proposed. Simulcast Video Coding is the parallel encoding of multiple independent video streams with different quality strategies (see Fig. 7). Each endpoint in the video conference chooses the most appropriate stream which it can process. This allows traditional endpoint, which doesn't support Scalable Video Coding technology to participate in the conference with the appropriate quality level (compatible with their bandwidth and CPU limitations).

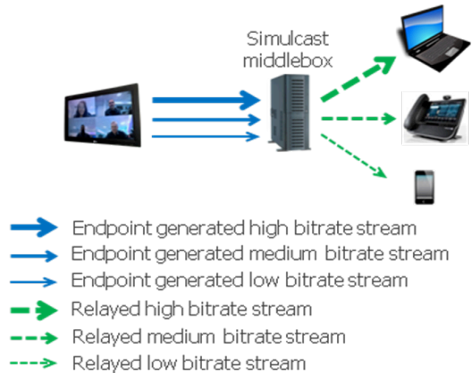

Fig. 7. Logic of Simulcast middlebox

In the Table 1 traditional, scalable and simulcast coding methods are compared by the following parameters:

- upstream bandwidth: the bandwidth needed to pass the client streams to the middlebox
- middlebox processing load: the amount of computations, the middlebox needs to execute in order to prepare the resulting streams for the clients
- downstream bandwidth: the bandwidth needed to pass the streams, prepared by the middlebox, to the client
- interoperability: to which extent the standard clients, which don't support a given technology, are able to receive the streams, coded by using this technology

Table 1. Comparison of coding methods

|  | Upstream bandwidth | Middlebox processing load | Downstream bandwidth | Inter operability |
|---|---|---|---|---|
| **Standard coding** | Low | High | Low | High |
| **SVC** | Low (with small overhead) | Low | Low (with small overhead) | Low (except base layer) |
| **Simulcast** | High | Low | Low | High |

### 3.3 Codecs

Several codecs are used in modern IP video conferencing. They are divided into several families.

*3.3.1 H.264/H.265*

H.264 and H.265 are the codecs standardized by ITU-T and ISO.

H.264 Advanced Video Coding (AVC) [2] was standardized in 2003. It provides a standard (non scalable) encoding with different quality levels, associated with the different sets of constraints imposed by decoder performance. The level defines the maximum picture resolution, frame rate and bitrate, that a decoder may use.

The H.264 standard was designed to be used by a wide variety of video applications, such as video conferencing, mobile video and high definition broadcast. Different types of target applications are addressed by profiles, which represent sets of coding tools and algorithms, used by the specific application (independently from the levels). Videoconferencing is typically based on the so called Baseline Profile (BP) or Constrained Baseline Profile (CBP).

H.264 is widely accepted in all areas and implemented in both hardware and software. H.264 is protected by a group of patents, which are managed by a holding of patent holders called "MPEG-LA".

H.264 Scalable Video Coding (SVC) [3] was standardized in 2007 and provides H.264 implementation of Scalable Video Coding.

H.265 High Efficiency Video Coding (HEVC) [4] is a successor to H.264 AVC. It was standardized in 2013. H.265 generally doubles the data compression ratio compared to H.264 AVC at the same level of video quality or, if used at the same bitrate, it substantially improves quality. H.265 SHVC (HEVC Scalability Extension) provides implementation of Scalable Video Coding technique and was published in 2015.

*3.3.2 VP8/VP9/VP10*

VP8, VP9 and VP10 are owned by Google. In 2010 Google exposed the source code of VP8 [5] under a 3-clause BSD license and the VP8 bitstream format is published by IETF. VP8 only supports temporal scalability.

VP9 [6] is a successor of VP8. It is also open and royalty free. Its bitstream description was published by IETF in 2013. The main improvement over VP8 is close to that of H.265 vs. H.264 – roughly half of bitrate is needed to deliver the same video quality. Scalable Video Coding version of VP9 is under development. VP10 is the latest evolution of this family which is in the early stage of development.

*3.3.3 NETVC*

In 2015 an Internet Video Codec (NETVC) working group was created at IETF with the target to produce a high-quality video codec meeting following requirements:

- competitive in terms of performance with best-of-the-breed existing codecs
- optimized for use in interactive web applications
- patent and royalty free allowing wide implementation and deployment

For the time being two codecs were submitted to IETF NETVC group: Daala and Thor.

Daala [7] is a free open source codec under development by Xiph.Org Foundation. The codec is developed based on the new principles compared to existing widely adopted codecs, which will allow avoiding patent infringement. The ideas of the codec are covered by some patents which are freely licensed to everybody.

Thor [8] is a free open source codec under development by Cisco. The target is to propose a codec of moderate complexity to allow real-time implementation in software on common hardware, as well as new hardware designs. Thor is based on technologies used in currently widespread standards.

*3.3.4 Alliance for Open Media*

The Alliance for Open Media [37] was founded by leading media related companies in 2015. The first target of the alliance is developing a new open royalty-free video codec specification and open-source implementation. VP10, Daala and Thor are considered for the development of this new codec.

*3.3.5 Other codecs*

In existing deployments, a wide variety of other codecs is present such as legacy codecs (H.261, H.263, …) and non-standard private codecs (Microsoft RTV, …).

## 4 Video processing

Given the set of video streams produced by conference participants' endpoints, the conferencing software needs to apply necessary processing in order to guarantee that all participants receive the streams that they are able to render. Processing generally consists of two parts: video presentation and video transformation.

**4.1 Video presentation**

Video presentation combines the streams, generated by the participants, in order to propose necessary user experience to stream recipients. Video presentation takes place in the middlebox or in the recipient endpoint. Today several types of user experience can be offered, depending on the capabilities of conferencing hardware/software and on the type of the conference considered.

*4.1.1 Video mixing (Continuous presence)*

Continuous presence mode is the most common method used in virtual meetings. Usually in this mode the screen is split into one large and several smaller surrounding windows. The conferencing software sends the video of the current speaker to the large window and other participants to the small ones. It's also possible to use equal windows for all the participants. If the number of participants is too large to show them all, only the latest speakers are displayed.

*4.1.2 Video switching*

Voice switch mode has only one window to which the conferencing software switches the current speaker.

*4.1.3 Lecturer mode*

In lecturer mode, the lecturer is shown all the time in the sole window. This mode is used for lectures and presentations.

*4.1.4 Chair mode*

In chair mode, a human moderator manually controls who "owns the floor", that is who can speak and who is shown on the screen at any given time.

*4.1.5 Augmented reality*

In augmented reality mode, participants are put into virtual meeting room by replacing background, and some additional video effects (like manipulation of the objects) can be added. At the present time, this mode is considered as experimental and is not widely implemented in the commercial products.

**4.2 Video transformation**

Video transformation is needed in order to adjust streams to the receivers' needs in the case when they can't be accepted in the original form. Video conferencing middlebox can process video streams on the level of stream content or on the level of stream packets, as explained hereafter.

*4.2.1 Content transformation*

Content processing means that some changes are introduced to the content of the video stream. This type of processing requires two-step. The first step is decoding, that is the stream encoded in original format is transferred to an uncompressed format. The second step is re-encoding, that is the uncompressed stream is encoded in a new format, taking into account the necessary changes, which should be introduced to the stream. During this two-step process the quality of the video stream suffers as lossy codecs are used in videoconferencing.

Content processing includes:

- TransCoding: change codec format in the case when the consumer doesn't use the same codec as the producer
- TransScaling: change the video frame size in the case when the receiver can't process big frames
- TransFrameRating: decrease video frame rate in the case when the receiver can't process too frequent frames
- TransBitRating: decrease the video codec bitrate which is the result of a decreased picture quality (i.e. bit per pixel).

The last three techniques are used in the case of scarce receiver resources or available network bandwidth.

*4.2.2 Packet transformation*

Packet processing implies that the middlebox processes IP packets without decoding/encoding the stream. Such processing contains:

- Packet filtering
- Packet forwarding
- Packet header correction

Packet processing mode can be used only when all the endpoints are compatible in the terms of codecs and their configuration, as packet processing server only decides which streams should be sent to each participant, filters necessary packets, changes header if necessary and forwards the packets to the respective endpoints. This technique is used extensively in Selective Forwarding Middlebox (which is described in section 6.3).

**5 Protocols**

The protocols, used in videoconferencing systems, can be considered on three levels:

- Media plane

- Signaling plane
- Media resources control

**5.1 Media plane**

The media plane (or data plane) consists of a set of protocols, used for transportation of audio and video streams on an IP packet network. Video conferences use Real-time Transport Protocol (RTP) [9] as a means of delivery of audio and video between endpoints and middleboxes. RTP is an application layer protocol based on UDP. The specificity of real-time audio and video favors the speed of delivery of the packets over reliability. TCP is generally used to get a reliable transfer, and is used for video streaming for instance. However, retransmission mechanisms of TCP introduce additional delay and jitter, which significantly lowers the quality of real-time interactive media sessions. That's why UDP is generally preferred for video conferencing. However, UDP itself is not sufficient and RTP provides facilities for jitter compensation and detection of out of sequence arrival in data, which are common during transmissions on an IP network. This is ensured by the addition of timestamps and sequence numbers.

RTP is used with its pair protocol "RTP Control Protocol" (RTCP), which provides statistics on the Quality of Service (QoS) and control information for an RTP session.

To overcome potential packet loss, some advanced technologies may be used in addition to RTP.

Packet Loss Concealment (PLC) is a technique to mask the fact that some video stream packets are lost, corrupted or arrived too late to be rendered. PLC uses info from neighboring parts to the lost segment of the frame, and/or previous frames and future frames, in order to estimate the lost content.

Forward Error Correction (FEC) is a technique, which adds redundant information to the video stream that can be used in order to recover lost packets of the stream.

Packet Retransmission (RTX) is a technique for retransmitting lost RTP packets by the source. This approach has limited use as it adds end-to-end delay, which consists of the time to make a request for retransmission and the time for a retransmitted packet to be delivered to a destination. Good Quality of Experience (QoE) [30] prescribes maximum one way delay of 150 ms in order to provide good media quality. If the resulting delay, introduced by packet retransmission, is bigger – this technique can't be employed.

**5.2 Signaling plane**

The signaling plane contains the protocols, that negotiate the creation/modification/termination of calls between the endpoints and the middleboxes. There exist both standard as well as proprietary signaling protocols.

Session Initiation Protocol (SIP) [10] is a text based protocol standardized by IETF. Its design is close to HTTP. Nowadays SIP is the main standard protocol for new developments in video conferencing area. Session Description protocol (SDP) is used by SIP as a means to exchange media related information, like the list of supported codecs, etc.

H.323 [11] is a binary protocol standardized by ITU-T. Before the wide adoption of SIP, it was the main protocol for videoconferencing, so many existing videoconferencing installations are still based on H.323. Currently, a lot of equipment supports dual SIP/H.323 stack.

Jingle [12] is an extension of eXtensible Messaging and Presence Protocol (XMPP), originally developed for chat services, which provides peer-to-peer signaling for multimedia sessions. It was developed by Google and the XMPP Standards Foundation, and used in a list of products, first of all open sourced.

Many well known products use their own proprietary signaling, for example Skype.

Network Announcement (NETANN) [13] provides a user with a possibility to call a conference (i.e. a virtual meeting room) using SIP. NETANN provides a way to use standard SIP messages, which were initially designed to locate and call a user, in order to locate and invoke a conferencing service. NETANN covers only very basic functionalities, not allowing rich conferencing user experience. In the scope of the SIPPING working group, the IETF presented a more advanced conference framework, based on SIP and described in [14]. The framework defines the logical entities and terminology used for conferencing. By the way, it was stated that while some conference management requirements can be implemented with SIP, some can't be implemented. Hence additional means are needed, as presented in the next section.

**5.3 Media resources control**

Media resource signaling takes place between the application server and the media server functions, in order to provide advanced conference control, such as:

- in-conference user interaction, like playing a voice message to the participants

- managing sub-conferences
- recording
- modification of the volume of a participant
- muting a participant
- conference event reporting (new participant is added, etc.)

There have been efforts to standardize the centralized control of a video conference at the IETF XCON working group [17]. There has also been work to standardize the control of a media server at the IETF MEDIACTRL working group [18]. However, these approaches are not widely implemented in the commercial products.

Media Server Markup Language (MSML) [15] is a XML based language, which is used to control conferencing features, such as video layout and audio mixing, configure media streams, create sidebar conferences or personal mixes. MSML was described by IETF in 2010.

Media Server Control Markup Language (MSCML) [16] is another XML based language, which also provides features to manage a conference similar to MSML. MSCML was described by IETF in 2006. MSCML and NETANN are related languages, as MSCML is an extension of NETANN.

In both cases, the XML messages are exchanged using the SIP protocol. Both MSML and MSCML are royalty-free and well adopted by the industry.

There also exists a "JSR 309: Media Server Control API" [19], which exposes media server control concepts in a form of Java API.

Finally, it's worth noting that the IETF CLUE working group (ControLling mUltiple streams for tElepresence) [20] has been created to develop a standardized approach to control immersive telepresence systems (see section 7.1). The management of such a system is much more complex, with its large number of screens and cameras, and requires specific exchange of capabilities and layout of the meeting rooms (e.g. spatial relationships of cameras, displays and microphones).

## 5.4 Protocols security

Network connections, used by video conferences, in particular when used over the Internet, are vulnerable to different security threats. Besides well known threats, such as Denial of Service (DoS) attacks, there exist specific security attacks, associated with the protocols listed above.

Signaling/media eavesdropping: this is an interception of signaling messages or media packets and extracting their content, which allows an attacker to know who is in communication with whom and what they are talking about.

Signaling/media spoofing: if signaling or media messages are intercepted, and their content is extracted, it becomes possible for an attacker to substitute the original participant with another one and to send false media streams to the conference participants — with eventually offending content for instance.

In order to avoid these problems, the video conferencing connections should be secured with the following requirements:

- authentication: in order to insure that each participant know with whom exactly it talks to
- encryption: packets should be encrypted in order to make it impossible to read the contents
- integrity: packets should be resistant to any changes introduced by an attacker

Secure version of SIP is called SIPS and consists in SIP over TLS, which gives SIP all necessary characteristics.

The media plane of video conferences can't be protected by TLS because it uses UDP (while TLS is based on TCP). The Secure RTP (SRTP) protocol has thus been defined at IETF [21]. SRTP requires participants to exchange cryptographical keys, and several mechanisms have been proposed: ZRTP [22], MIKEY [23], SDES [24] and DTLS-SRTP [25].

## 5.5 Protocols network border traversal

Enterprise network borders are normally protected by a firewall, which provides Network Address Translation (NAT) and traffic filtering functions. Firewalls pose certain problems for both video conferencing signaling and media traffic, which can't bypass the network border.

Special protocols were introduced for traversing a NAT device. Session Traversal Utilities for NAT (STUN) protocol [26] is used by an endpoint to determine public IP address and port allocated to it by a NAT device. Traversal Using Relays around NAT (TURN) protocol [27] is used, when knowledge of a public IP address is not sufficient in order to traverse a NAT device, and an external server, which relays media streams is needed. TURN allows an endpoint to control the operation of such a relay.

STUN and TURN are combined in Interactive Connectivity Establishment (ICE) protocol [28], which enables the procedure of discovery and exchange of the info, which is needed to establish a media connection in the presence of NAT.

The filtering function for video conferencing is usually executed by a Session Border Controller (SBC). SBC is deployed at the network border and provides firewall function for both signaling and media traffic. SBC can provide a video transformation function (see section 4.2) as well.

### 5.6 Protocols compatibility

One of the main problems of the industry has been incompatibility of the solutions of different vendors for a long time. Signaling protocols as well as configuration of codecs in use were implemented differently within the limits of the respective standards.

Several industry organizations, like International Multimedia Telecommunications Consortium (IMTC) [34] and International IP Interconnection Forum (i3forum) [35] unite leading video conferencing players in order to eliminate discrepancies in protocols implementation.

## 6 Types of media servers

Media servers, used in videoconferencing, can be of several types, reflecting their design and functionality.

### 6.1 Multipoint Control Unit

Multipoint Control Unit (MCU) [29] is a hardware appliance or software component, which provides both video presentation and video transformation functionalities. It means that it is capable of providing any type of stream presentation (mixing, switching, other types of presentation, requiring stream content processing) as well as stream modification (trans-coding, etc). An MCU can also be referred to as a "conference bridge".

The standard MCU functionality is to decode all incoming media streams, compose a particular stream for each conference participant and encode these streams to send them through the network. MCU performance is counted in ports, one port being capable to receive one video stream.

Traditionally, hardware MCUs utilize Digital Signal Processors (DSP) in order to make operations over video streams more efficient. Evidently, hardware MCUs are not easily scalable, as if you need more ports you need to buy more physical instances of the hardware.

A Software MCU is a software component, which is deployed on standard general purpose server hardware. Latest advancements in CPU technologies and associated programming libraries, such as the Intel Integrated Performance Primitives (Intel IPP) library [36], have made it possible for general purpose CPUs to be used for video processing. Contrary to a hardware MCU, Software MCU provides efficient scalability and flexibility. Software MCUs can be deployed in the cloud with the access to the exact amount of hardware resources, which are needed. That makes the operation of adding and removing of MCU ports very simple. Furthermore, Software MCUs benefit from easy operations of update and upgrade compared to hardware MCUs.

Hardware and software MCU logic is depicted in Fig. 8.

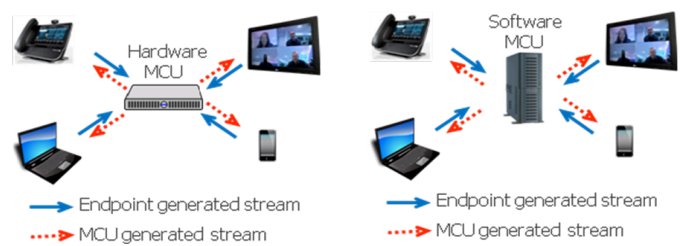

Fig. 8. Logic of MCU

### 6.2 Video Gateway

A video gateway is a hardware appliance or a software component, which provides video transformation functions, being a mediator between two incompatible video conferencing systems (e.g. from two different vendors). The incompatibility may be on signaling and/or media level. Such a way video gateway can provide:

- signaling: connecting different protocols (e.g. H.323/SIP), or aligning different flavors of SIP
- transport: changing the transport protocol (TCP/UDP)
- media: trans-coding, adjusting modes of the same codec
- security: interworking between secure side (SRTP) and insecure side (RTP)

Video gateway can also be used for some additional services, which are based on information passing through it, for example for conference recording.

### 6.3 Selective Forwarding Middlebox

Selective Forwarding Middlebox (SFM) [29] is a software component, which relays the received video streams to the different conference participants. For that purpose SFM can apply packet transformation functionalities. As SFM doesn't provide content processing, it doesn't consume a lot of CPU cycles as compared to MCU.

Often SFM may receive different resolutions of the stream from one conference participant based on SVC or simulcast technology, so it should apply some logic in order to decide which resolution to send to each recipient. This logic can use the physical characteristics of the recipient in order to choose appropriate resolution. It may also be used to detect active speaker among conference participants in order to send her stream with better resolution than others.

SFM can be based on:

- standard video coding, in the case when the endpoints are all supporting this encoding
- Scalable Video Coding
- Simulcast Video Coding

In the Fig. 9. an SFM based conference with three participants is depicted. All the three participants send two levels of SVC coding. Two non-active participants, using devices with relatively small screens, receive one high quality stream with the active speaker and one low quality stream with the other non-active participant. Active speaker, using a device with a big screen, is able to benefit from receiving all the streams in high quality level (even if these are two stream of non speaking participants).

## 7 Types of clients

### 7.1 Immersive telepresence

Immersive telepresence is the most advanced (and also the most expensive) form of the video conferencing, providing the users with the experience that all the conference participants are located in the same room. Purposely designed conferencing rooms equipped with several large screens and a variety of cameras are usually used for immersive telepresence.

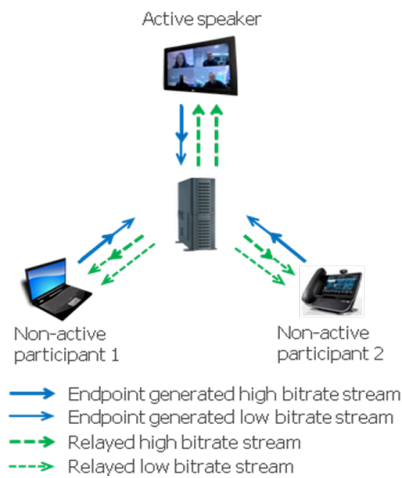

Fig. 9. Logic of SFM based on Scalable Video Coding

### 7.2 Hardware clients

Video conferencing hardware clients are dedicated appliance, that can be used in meeting rooms or at working desks in order to participate in a conference. Usually such a client comprises a webcam and a loudspeaker. Also it can be equipped with a screen or it can use a 3rd party screen.

### 7.3 Video desk phones

Modern desk phones have big screens, which can host an image produced by a videoconference. Integrated or external webcam is used with such a solution.

### 7.4 PC clients

PC clients are installed on user's computers and use webcam and screen of this computer. The quality of video stream which is produced/consumed by such a client depends on CPU of the hosting computer.

### 7.5 Browser clients

The software clients, which don't need installation, but are instead downloaded from a web site and ready to be used immediately. This category is now gaining traction especially after the introduction of the WebRTC technology. WebRTC standardizes interaction between conferencing web application and web browser, based on JavaScript API [32]. Also it provides a full media stack (protocols, codecs, …) [33].

### 7.6 Mobile clients

With the rise of smart phones and tablets, mobile devices have become a popular platform for video conferencing clients. Taking into account that such clients are battery powered, the codec implementation shout pay special attention to energy consumption.

## 8 Topologies

### 8.1 Dedicated on-premises

Videoconference systems are traditionally implemented as a hardware appliance, deployed in the LAN of an enterprise, or as a software component, installed on one or more servers in the data center of an enterprise.

### 8.2 Hosted

A Hosted deployment means that hardware appliance is physically located in the data center of a service provider and operated by its IT team (see Fig. 10). Deployed hardware can be used by several clients or it can be locked to a single client in order to provide more security. Hosted deployment should not be confused with Cloud, as the former still uses hardware appliances, and, as such, doesn't provide easy scalability.

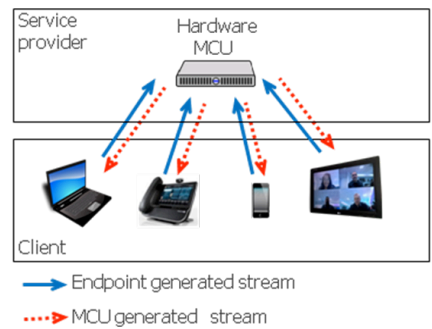

Fig. 10. Hosted hardware MCU

### 8.3 Cloud

A Cloud deployment refers to a software component in the form of a virtual machine, which is physically located in the data center of a service provider (see Fig. 11). All types of middleboxes (MCU, SFM, gateway) can be deployed in the Cloud. Service is offered either as a subscription (customer pays for each registered user) or on a usage basis (cost per port per minute).

A Cloud deployment has, as defined in [31], the following properties:

- On-demand self-service
- Broad network access
- Resource pooling
- Rapid elasticity

- Measured service

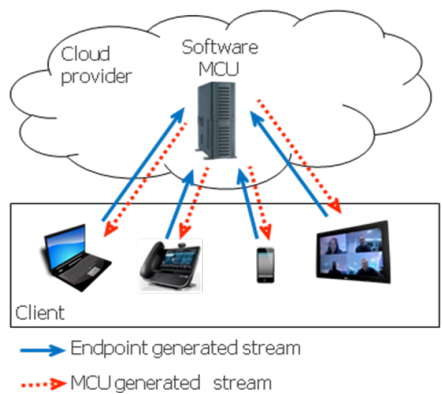

Fig. 11. Software MCU in the Cloud

**8.4 Enterprise Desktop Grid**

An Enterprise Desktop Grid deployment means that video processing software component is deployed on the grid of usual office hardware, that is desktop and laptop PCs (see Fig. 12). The machines, which are not occupied with other activities, are selected to host videoconferencing processing tasks. All types of middleboxes (MCU, SFM, gateway) can be deployed on Enterprise Desktop Grid. This is an experimental type of deployment, which is under investigation [39].

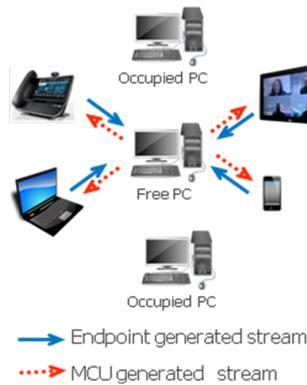

Fig. 12. Video conferencing deployed on Enterprise Desktop Grid

The different approaches to video conference deployments are compared in the Table 2. The following criteria are considered:

- Expenditure : how the conference service is paid for

- Support: who is responsible for day-by-day operations and administration

- Data: where and how the data, related to the conferencing solution (recorded video, logs, …), is stored and who can have access to it

- Elasticity: to which extent the conferencing solution can be scaled

Furthermore, there exist two special cases for the deployment of the transformation and presentation functionalities (see chapter 4).

Table 2. Comparison of deployment approaches

|  | Hosted | Cloud | On-premises hardware | On-premises software | Enterprise Desktop Grid |
|---|---|---|---|---|---|
| **Expenditure** | Operational (OPEX) | | Capital (CAPEX) | Capital (CAPEX), general purpose servers are used | Free, as already existing hardware is reused. |
| **Support** | Service provider | | Local IT | | |
| **Data** | Questionable when contract is stopped, potentially can be accessed by 3$^{rd}$ parties | | Always accessible by owner, not accessible by 3$^{rd}$ parties | | |
| **Elasticity** | Limited by hardware MCU | Not limited | Limited by hardware MCU | Limited by servers in datacenter | Depends on load of the grid |

### 8.5 Conference endpoints

The video presentation logic can be located in the conference endpoints. In this case a conference is organized without a middlebox, in a peer-to-peer fashion (see Fig. 13).

Advantages:

- No additional resources are required

Disadvantages:

- It is not possible to support many participants, as any endpoint is required to keep a separate video connection with all the other participants.

- It is not possible to support heterogeneous endpoints: all the participants must use the same type of client software or hardware so that codecs and their parameters (resolution, frame rate, etc) match. There is actually no gateway functionality in peer-to-peer topology, by construction.

- Absence of enterprise conferencing features: enterprise use cases often require more advanced features like conference recording, possibility to add/remove participants by a moderator, etc., which is difficult to offer in this distributed environment.

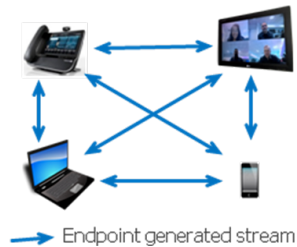

Fig. 13. Full mesh conference topology

In order to optimize video streams, multicast technology can be utilized. Two types of multicast technologies are available: Application Level Multicast (ALM), where the distribution of flows towards the different receivers is performed by endpoints themselves (i.e. at the application layer) (see Fig. 14), or directly by the network (IP Multicast).

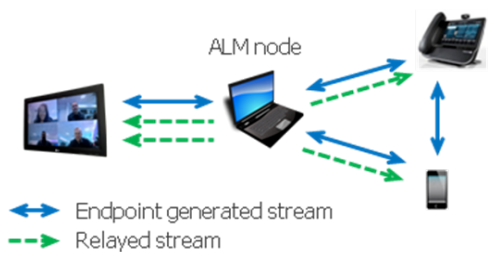

Fig. 14. Conference topology based on Application Level Multicast

IP multicast has, however, not gained traction, because:

- Internet Service Providers usually don't offer IP multicast services, by reason of management complexity, scalability concerns and security risks, associated with this technology. As a consequence, this technology is not available when the videoconference takes place over the Internet (involving participants from different networks).
- In an intranet context, multicast is much widely available. However, in the enterprise environment, additional services are needed, which can only be provided by a middlebox. ALM is thus more natural in this context.

Selective Forwarding Middleboxes can also be used in the context, when video processing is integrated in the endpoints:

Advantages:

- Often the most cost-effective choice

Disadvantages:

- No trans-coding of different video and audio codecs
- Limited number of participants (usually less than 6)
- Bandwidth requirements aggregate to increase demand on the organizer's network

**8.5 Hybrid topologies**

Different types of hybrid topologies are possible. Particularly on-premises + cloud deployments are popular nowadays. If a company has got through mergers and acquisitions a fleet of several MCUs of different vendors, it can use a cloud multi-vendor interoperability service, which allows using these MCUs in the same conference. Or a company can decide to physically protect its data in the form of recordings of the meetings, and place the recording and storage server in their data center while the conferences consume resources in the cloud.

**9 Current trends**

Currently, several trends of video conferencing evolution can be identified.

In the codec area, there is a significant move towards open royalty-free codecs, which is supported by a great number of industry players through "Alliance for Open Media" and IETF NETVC working group. Further increase of resolution from today's standards HD (1280×720) and Full HD (1920×1080) towards 4k (4096×2180) also gains traction for big screens.

Pure software technologies, on which both server infrastructure and clients are based, are extensively developed nowadays.

On the server side they allow virtualization, which is a necessary requirement in order to place video conferencing to the cloud and expose it in the form of VCaaS (Video Conferencing as a Service).

At the same time, we can wait evolutions of a Fog approach [38], which moves processing burden from the cloud to the resources in proximity of the end users, for example to the network edge devices. The Fog technique can potentially spare needed WAN bandwidth as well as reduce end-to-end delay.

On the client side, software technologies allow using commodity hardware (PC screens, integrated cameras and microphones), instead of purchasing expensive one-purpose videoconferencing endpoints. This trend is especially important with wide adoption of mobile clients (smart phones, tablets) as video conferencing endpoints for the workers, which are not attached to a fixed working place.

WebRTC technology becomes very popular, as it enables browser based clients, which are accessed by the users as standard web pages. This removes the necessity of client installation, significantly improving end user experience. Being oriented to numerous web developers, the technology is very popular and benefits from wide support of the community.

In today's world, video conferencing is less considered as a stand-alone technology and more as a part of "collaboration", which is a wider notion, comprising different means of meeting organization like instant messaging and screen sharing as well as integration of video conferencing functionality to business and vertical applications, which prescribes its efficient usage in the context of business and industrial processes of the organization.